\documentstyle[aps,prbbib,twocolumn,epsf]{revtex}

\begin{document}
\topmargin -1.4cm

\draft
\title{Kondo resonance in a multi-probe quantum dot}

\author{Qing-feng Sun and Hong Guo}

\address{Center for the Physics of Materials and Department
of Physics, McGill University,   Montreal, PQ, Canada H3A 2T8}

\maketitle

\begin{abstract}
We present a theoretical analysis of a possible route for directly 
detecting Kondo resonances in local density of states (LDOS) 
of an interacting quantum dot. By very weakly couple a third 
and/or a fourth lead to a two-probe quantum dot and measuring 
differential conductance through these extra links, we show that 
Kondo peaks directly map onto the differential conductance measured
from the third link. We analyze the conditions by which this detection 
of Kondo peaks in LDOS is possible. 
\end{abstract}

The Kondo effect is a prototypical many-body correlation 
effect in condensed matter involving the interaction 
between a localized spin and free electrons. At low temperatures 
and in the Kondo regime, the entire system---localized spin
plus free electrons, forms a spin singlet state and the localized 
spin is screened. As a result, a very narrow Kondo resonance 
peak arises in the local density of states (LDOS) at the
chemical potential of the free electrons. The Kondo effect was
first discovered in metals containing a small amount of 
magnetic impurities. More recently it was observed in 
semiconductor quantum dots (QD)\cite{ref1,ref2} which has
generated a great deal of theoretical and experimental 
interests because it provides rich understanding 
to many-body phenomena at the mesoscopic scale.

Experimental investigations of Kondo phenomenon in semiconductor QD were
mainly through two observations\cite{ref1,ref2}. First, for cases of 
QD confining an odd number of electrons, the differential conductance 
$dI/dV$ is measured as a function of a gate voltage $V_g$, here $I$ 
is the two-probe tunneling current and $V$ the bias voltage. 
It was found that $dI/dV$ in the Coulomb blockade region is 
enhanced\cite{ref1,ref2} due to the Kondo effect. Second, there is a 
peak at bias $V=0$ in the $dI/dV$ versus $V$ curve, and this peak splits 
into two when there is a magnetic field\cite{ref1}. Although $dI/dV$
gives a measure of LDOS of the QD in linear response, to the best of our
knowledge the comprehensive shape of the LDOS of the QD in the Kondo 
regime, namely the one (or a few) narrow Kondo peak on top of the 
``shoulder'' of the broad peak corresponding to a intradot level, 
has so far not been directly detected in any experiment.

The outstanding features in the QD Kondo phenomenon are most prominent and
plentiful in LDOS than in the tunneling current $I$ and its associated
differential conductance $dI/dV$. For example, for a QD coupled to a normal 
lead and a superconducting lead (a N-QD-S device), three Kondo peaks 
arise in LDOS at the chemical potential of the normal lead and at 
the superconducting gap ($\pm\Delta$) respectively. On the other hand,
the tunneling current hardly vary at all despite the presence of
these Kondo peaks in LDOS\cite{ref3}. As another example, in an asymmetric N-QD-S 
device under a finite on-site Coulomb interaction and a large 
superconducting gap, four Kondo peaks emerge in LDOS\cite{ref4}. However, while the 
current is enhanced due to these features, it does not show clear
characteristics of the narrow Kondo resonances. Therefore, it is
extremely useful to be able to directly detect the narrow Kondo 
resonance in the LDOS.

Given the importance of the physics of Kondo phenomenon in 
mesoscopic systems and the extensive investigations in both 
theory and experiments, it is indeed surprising to see the 
lack of direct observation of the Kondo resonance peaks 
in the QD LDOS\cite{ref5}. It is the purpose of this
communication to present a theoretical analysis of a 
possible route for solving this problem. Because, as
discussed above, that two-probe tunneling differential
conductance $dI/dV$ does not reveal the narrow Kondo peaks
in LDOS, we will investigate a new approach by which
one or two extra leads are used to probe the QD. When
conditions are controlled correctly, we show that 
the LDOS (including the narrow Kondo peaks) will directly 
map onto the current measured at the extra leads thereby 
providing a direct measurements of the narrow Kondo peaks 
elusive so far.

To begin, let's consider the hypothetical device consisting
of a QD coupled to {\it four} leads fabricated by a split
gate technique, as shown in the inset of Fig.2, in a 
two-dimensional electron gas (2DEG). Here, leads 1 and 3 
plus the QD form a typical two-probe QD device for which we
assume as having a Kondo regime at low temperature, so that
there are some Kondo resonances in the LDOS which is 
our target of measurement. Leads 2 and 4 are assumed to very 
weakly couple with the QD, much weaker than that of leads 1 
and 3. The quantification of these statements will be made below.
Our hope is to probe the QD Kondo physics through lead 2. 
The purpose of the bias on lead 4, $V_4$, is to provide a 
voltage opposite in sign to that of $V_2$, so as to compensate 
the intradot energy altered by bias $V_2$. Our results suggest 
that when conditions are right, the differential conductance 
$dI_2/dV_2$ versus its terminal bias $V_2$, gives an 
excellent measurement to the LDOS as a function of energy 
$\epsilon$, thereby allowing us to observe the Kondo peaks 
directly.

There are so far some literature attempting to measure the Kondo 
resonance in LDOS\cite{ref5,ref6,ref7}. They use a scanning 
tunneling microscope (STM) to obtain spectroscopic data 
on individual magnetic impurities deposited onto the host 
metal. However, due to quantum interference between the $d$ 
and/or $f$ orbitals of the magnetic impurity and the 
continuum conduction electron channels, so far only Fano 
like resonances were obtained in the spectrum rather than
the expected Lorentzian shape of the Kondo peaks. In contrast, 
our present hypothetical experiment will attempt to directly
obtain the original shape of the Kondo resonance in LDOS. 
The essential difference in our device as compared to 
those studied before are the following. In the STM 
experiment\cite{ref7,ref8}, there exists electron transitions between 
the host metal, which provides conduction or free 
electrons, and the probing STM tip, such transitions cannot 
be avoided. In our device, there does not exist
direct tunneling between the probe terminal lead 2
and the conduction electron channels lead 1 and 3. This 
important difference allows us to observe the original shape 
of LDOS including the Kondo resonances.

Our device is described by the following Hamiltonian,
$$ H=\sum\limits_{\alpha,k,\sigma}
\epsilon_{\alpha k }a_{\alpha k \sigma}^{\dagger}
a_{\alpha k \sigma}
+\sum\limits_{\sigma} \epsilon_{d\sigma}d_{\sigma}^{\dagger}d_{\sigma}
+Ud^{\dagger}_{\uparrow}d_{\uparrow}d^{\dagger}_{\downarrow}d_{\downarrow}
$$
\begin{equation}
+ \sum\limits_{\alpha,k,\sigma} \left(
v_{\alpha k}a_{\alpha k \sigma}^{\dagger}d_{\sigma} + H.c.
\right) \hspace{1mm},
\end{equation}
where $a_{\alpha k \sigma}^{\dagger}$($a_{\alpha k \sigma}$) 
($\alpha=1,2,3,4$) and $d_{\alpha}^{\dagger}$
($d_{\alpha}$) are creation (annihilation) operators in the
lead $\alpha$ and the QD, respectively. The QD includes a 
single energy level, but having spin index $\sigma$ and 
intradot Coulomb interaction $U$. To account for a possible
magnetic field, we allow $\epsilon_{d\uparrow}\neq
\epsilon_{d\downarrow}$. The last term describes 
the tunneling part of the Hamiltonian, with $v_{\alpha k}$ 
being the coupling matrix element. 

The current from lead $\alpha$ flowing into the QD can be 
expressed as\cite{ref9},
\begin{equation}
   I_{\alpha} = -2e Im \sum\limits_{\sigma} \int \frac{d\epsilon}{2\pi}
\Gamma_{\alpha} \left\{ f_{\alpha}(\epsilon)G^r_{\sigma}(\epsilon)
+\frac{1}{2} G^<_{\sigma}(\epsilon) \right\} ,
\label{iv1}
\end{equation}
where $\Gamma_{\alpha}(\epsilon)=2\pi \sum_k |v_{\alpha k}|^2
\delta(\epsilon-\epsilon_{\alpha k})$; $f_{\alpha}(\epsilon)$ is 
the Fermi distribution of lead $\alpha$; $G^r_{\sigma}(\epsilon)$ and
$G^<_{\sigma}(\epsilon)$ are the retarded and the Keldysh Green's
functions of the QD, they are the Fourier transformation of 
$G^{r,<}_{\sigma}(t)$, and 
$G^r_{\sigma}(t)\equiv -i \theta(t)<
\{ d_{\sigma}(t),d_{\sigma}^{\dagger}(0)\}>$,
$G^<_{\sigma}(t)\equiv  i<d_{\sigma}^{\dagger}(0)d_{\sigma}(t)>$.

Using the standard equation of motion technique and taking the 
familiar decoupling approximation\cite{ref11}, we have
solved $G^r_{\sigma}(\epsilon)$ to be
\begin{equation} 
G^r_{\sigma}(\epsilon)= \frac{1+U A_{\sigma} n_{\bar{\sigma}} }
{\epsilon-\epsilon_{d\sigma}-\Sigma^{(0)}_{\sigma}
+U A_{\sigma} (\Sigma^{(a)}_{\bar{\sigma}} + \Sigma^{(b)}_{\bar{\sigma}}) }
\hspace{1mm},
\label{green1}
\end{equation}
where 
$
 A_{\sigma}({\epsilon})=[\epsilon-\epsilon_{d\sigma}-U
-\Sigma^{(0)}_{\sigma}-\Sigma^{(1)}_{\bar{\sigma}}-\Sigma^{(2)}_{\bar{\sigma}}]^{-1} 
$
and $\Sigma^{(0)}_{\sigma}=\sum_{k\alpha} 
|v_{\alpha k}|^2/(\epsilon-\epsilon_{\alpha k} +i0^+)$ 
is the lowest-order self-energy which is exactly the 
retarded self-energy for a noninteraction system;
$\Sigma^{(a)}_{\sigma}$, $\Sigma^{(b)}_{\sigma}$, 
$\Sigma^{(1)}_{\sigma}$, and $\Sigma^{(2)}_{\sigma}$ are 
the higher-order self-energies due to the intradot
Coulomb interaction and the tunneling coupling. 
These higher-order self-energies are derived to be: 
$\Sigma^{(a)}_{\sigma}=\sum_{k\alpha} |v_{\alpha k}|^2
f_{\alpha}(\epsilon_{\alpha k})/\epsilon_{\sigma}^{+}$;
$\Sigma^{(b)}_{\sigma}=\sum_{k\alpha} |v_{\alpha k}|^2
f_{\alpha}(\epsilon_{\alpha k})/\epsilon_{\sigma}^{-}$;
$\Sigma^{(1)}_{\sigma}=\sum_{k\alpha} 
|v_{\alpha k}|^2/\epsilon_{\sigma}^{+}$;
$\Sigma^{(2)}_{\sigma}=\sum_{k\alpha} 
|v_{\alpha k}|^2/\epsilon_{\sigma}^{-}$;
here $\epsilon_{\sigma}^{+}=\epsilon+
\epsilon_{\alpha k}-\epsilon_{d\sigma}
-\epsilon_{d\bar{\sigma}}-U+i0^+ $
and $\epsilon_{\sigma}^{-}=\epsilon-
\epsilon_{\alpha k}-\epsilon_{d\bar{\sigma}}
+\epsilon_{d\sigma}+i0^+ $. The quantity 
$n_{\bar{\sigma}}$ in Eq.(\ref{green1}) is the
intradot electron occupation number of state 
$\bar{\sigma}$, which needs to be calculated 
self-consistently\cite{ref12}. In the limit of having only
two leads, the above results reduces to that of Refs.\onlinecite{ref11,ref13}.

The Keldysh Green function $G^<_{\sigma}$, for interacting 
systems, can not be obtained from the equation of motion 
without introducing additional assumptions. We use the 
standard ansatz due to Ng\cite{ref14},
\begin{equation}
\Sigma^<_{\sigma}(\epsilon)
=-\sum\limits_{\alpha}\frac{\Gamma_{\alpha}f_{\alpha}(\epsilon)}
{\Gamma} (\Sigma^r_{\sigma}-\Sigma^a_{\sigma}) \ \ ,
\end{equation}
where $\Gamma=\sum_{\alpha}\Gamma_{\alpha}$. 
$\Sigma^<_{\sigma}$ and $\Sigma^r_{\sigma}$($\Sigma^a_{\sigma}$) 
are the lesser and retarded (advanced) self-energies of the 
interacting system. Then from the Keldysh equation 
$G^<_{\sigma}=G^r_{\sigma}\Sigma^<_{\sigma}G^a_{\sigma}$ and 
$G^a_{\sigma}= (G^r_{\sigma})^*$, $G^<_{\sigma}$ can be obtained
straightforwardly. With $G^r_{\sigma}$ and $G^<_{\sigma}$ solved, 
from Eq.(\ref{iv1}) the current can be obtained immediately:
\begin{equation}
I_{\alpha} = -2e \sum\limits_{\sigma, \alpha'} \int
\frac{d\epsilon}{2\pi} 
\frac{\Gamma_{\alpha}\Gamma_{\alpha'}}{\Gamma}
\left[ f_{\alpha}(\epsilon)-f_{\alpha'}(\epsilon)\right]
Im G_{\sigma}^r \ .
\end{equation}

In the numerical calculation, we make a few further 
simplifications: (i) we assume square bands of width $2W$, 
so that $\Gamma_{\alpha}(\epsilon)=\Gamma_{\alpha}
\theta(W-|\epsilon|)$, with $W=1000\gg max(k_B T,V_{\alpha}, 
\Gamma_{\alpha})$; (ii) we take the large $U$ limit 
$U\rightarrow \infty$; (iii) considering that the intradot 
level $\epsilon_{d\sigma}$ is affected by leads' bias 
voltage $V_{\alpha}$, we assume this effect to be
$\epsilon_{d\sigma}=\epsilon_{d\sigma}(0)
+\gamma_2 V_2 +\gamma_4 V_4$, with 
$\gamma_{\alpha} =C_{\alpha}/C$. Here $C_{\alpha}$ is 
the capacitance between lead ${\alpha}$ and the QD, 
and $C$ is the total capacitance of the QD; 
$\epsilon_{d\sigma}(0)$ is the location of the intradot 
energy level at $V_2=V_4=0$. We set $V_4=-V_2$ to offset 
the level change, so that 
\begin{equation}
\epsilon_{d\sigma}= \epsilon_{d\sigma}(0)+\gamma V_2 
\label{gamma1}
\end{equation}
where $\gamma=(C_2-C_4)/C$.

Our objective system is the QD plus leads 1 and 3, it is
recovered by setting $\Gamma_2=\Gamma_4=0$ so that leads
2 and 4 are decoupled from the QD. The dotted curves of 
Fig.1 shows the intradot LDOS of the objective system\cite{ref15}.
A broad peak at $\epsilon= -2$ is due to the intradot 
renormalized level. In nonequilibrium and at zero magnetic field 
($\epsilon_{d\uparrow}=\epsilon_{d\downarrow}$),
there exhibits two narrow Kondo resonance peaks at $\mu_1$ and 
$\mu_3$ in the LDOS. With a non-zero magnetic field
($\epsilon_{d\uparrow}\not=\epsilon_{d\downarrow}$) and in 
nonequilibrium, four narrow Kondo peaks emerge at
$\mu_{1/3}\pm\Delta\epsilon$ where
$\Delta\epsilon=\epsilon_{d\downarrow}-\epsilon_{d\uparrow}$ 
is the level difference. These familiar characters of 
LDOS have been known\cite{ref13} in theory, our task to 
``experimentally'' measure them.
                        
Let's now turn on a non-zero $\Gamma_2$ and $\Gamma_4$. Note
that these couplings must be greatly weaker than those 
of leads 1 and 3, {\it i.e.} $\Gamma_2, \Gamma_4\ll \Gamma_1,\Gamma_3$, 
so that they do not affect the QD significantly. We first consider the 
$\gamma=0$ case for which there is a complete compensation of 
$V_2$ by $V_4$ so that the level $\epsilon_{d\sigma}$ does 
not change with $V_2$ by Eq.(\ref{gamma1}). The differential conductance 
$dI_2/dV_2$ versus $V_2$ is shown in Fig.1 by the solid curves, and they
overlap almost identically with the dotted curves of the two-probe
LDOS (the objective system), so that the dotted curves can not be seen
in the fig.1. In other words, the $dI_2/dV_2$-$V_2$ data 
and the $LDOS$-$\epsilon$ data of the objective system map into each 
other essentially perfectly. To see better the comparison, data in the 
vicinity of Kondo peaks for zero magnetic field are shown in the 
inset of Fig.1. Although the Kondo peaks of $dI_2/dV_2$ 
is slightly lower than those of LDOS, they not only agree in their position 
but also in the Lorentzian shape which is a very important characteristic 
of Kondo phenomenon.

Why does $dI_2/dV_2$ versus $V_2$ give such an excellent mapping of the 
original LDOS of the objective system ?  First, lead 2 is very weakly 
coupled to the QD so that the original QD LDOS is not significantly 
affected by it. Second, because the QD is coupled to leads 1 and 3 in much
stronger way, {\it e.g.} $\Gamma_1,\Gamma_3\gg \Gamma_2$, resonance
tunneling from lead 2 to leads $1,3$ can not occur with any substantial
probability. Therefore, an incident electron with energy $\epsilon$ from 
lead 2 has a probability of tunneling into the QD that is given by  
the intradot LDOS($\epsilon$), leading to the excellent agreement 
between $dI_2/dV_2$ and the LDOS. We conclude that the LDOS versus
energy $\epsilon$ can be obtained by measuring $dI_2/dV_2$ versus $V_2$.

It is worth mentioning that although we have assumed symmetric couplings
between leads 1,3 to the QD, $\Gamma_1=\Gamma_3$, and assumed a large e-e 
interaction $U\rightarrow \infty$, it is straightforward to confirm, as we 
did, that if $\Gamma_1\not=\Gamma_3$ and $U$ is finite, our results are still 
tenable. In fact, these parameters of the objective system only affect its
LDOS, they do not destroy the excellent agreement between the signal
$dI_2/dV_2$ and the LDOS. In addition, the physics dictating this excellent
agreement is independent of what theoretical methods one uses to derive
the Green's functions $G^r_{\sigma}(\epsilon)$. In other words, if one
uses another method to solve $G^r_{\sigma}(\epsilon)$ rather than the
equation of motion method we used here\cite{ref13}, or even if
one gives an arbitrary LDOS of the target system, our proposed detection
technique can still give the excellent agreement between $dI_2/dV_2$ and 
the LDOS.

Other important issues concerning our proposal are the ranges of parameters
associated with leads 2 and 4 which we use to probe the LDOS. If the 
resistance provided by lead 1 to the QD is a typical $10K\Omega$, then if 
lead 3 couples $1000$ times weaker, its contact to QD will have a resistance 
of $10M\Omega$ which is experimentally not difficult. When voltages on
leads 2 and 4 do not exactly compensate, {\it i.e.} when $\gamma\neq 0$, 
the intradot level $\epsilon_{d\sigma}$ will change with $V_2$ according
to Eq.(\ref{gamma1}), hence features in LDOS will be changed which affect
the proposed measurement of the Kondo peaks by $dI_2/dV_2$. Our investigation
on this problem shows that this is actually a weak effect on the Kondo
resonances, as shown in Fig.2 where $dI_2/dV_2$ is plotted against
$V_2$ at several values of $\gamma$. The background differential
conductance does change with $\gamma$. 
However, the important result is that the narrow Kondo peaks still 
keep the original shape, and their locations do not vary at 
all as shown in Fig.2. Even when $\gamma=0.5$ or larger, these Kondo 
characters remain. Therefore, we believe the condition on parameter
$\gamma$ is not strict for our proposal to work.
It should be stressed that if $\gamma$ is not very large, {\it e.g.}
$\gamma=0.05$, the broad peak which corresponds to the intradot
renormalized level, only shifts slightly but retaining its line shape
(dashed line of Fig.1). So the comprehensive shape of LDOS in the
Kondo regime, the one (or a few) narrow Kondo peak on top of 
the shoulder of the broad peak, can be detected.

Let's estimate the value of $\gamma$ which, of course, is less than unity. 
The total capacitance $C$ includes each terminal capacitance $C_{\alpha}$
($\alpha=1,2,3,4$), it also includes, perhaps, some gate capacitances $C_g$. 
A gate is a negatively biased metal deposited above the 2DEG while the leads' 
terminal is in the 2DEG and coupled to QD by tunnel barriers. In general, 
$C_{\alpha}/C_{\alpha'}$ is approximatively proportional to 
$\Gamma_{\alpha}/\Gamma_{\alpha'}$, and $C_g$ is larger than 
$C_{\alpha}$\cite{ref17}. Because $\Gamma_2 \ll \Gamma_1$ and $\Gamma_3$
(in Fig.1 and 2, $\Gamma_1= \Gamma_3 =1000 \Gamma_2$), 
a conservative estimate is that $C_1$ and $C_3$ should be larger than 
$10C_2$. Then, even if we neglect $C_g$ in the total $C$ and even if we 
completely remove the compensating terminal 4, we still have
$\gamma=C_2/C <0.05$. Moreover, if the compensating lead 4 is there to
cancel the effect of the probe terminal 2 to some extent, we estimate 
$|\gamma|$ should be less than $0.01$. Hence, by results of Fig.2, such a 
small $\gamma$ will not cause any trouble to our proposed measurements of 
the Kondo peaks.

Next, we discuss the range of temperature $T$ where the excellent 
agreement of $dI_2/dV_2$ and LDOS can be kept. We fix $\gamma=0.05$,
as discussed above, this value of $\gamma$ is easily realized even without
the compensating lead 4. Therefore we set, in our theory,
$\Gamma_4=0$ and $C_4=0$, and we propose to replace lead 4 by a gate 
to control the intradot level position (see inset of Fig.3).
The signal $dI_2/dV_2$ versus $V_2$ is shown in Fig.3 for three temperatures. 
At low temperature, there are four Kondo peaks at 
$\mu_{1/3}\pm\Delta\epsilon$ in $dI_2/dV_2$, they are the mapping of
the four Kondo peaks in $LDOS(\epsilon)$ which have been discussed above. 
Increasing temperature causes the Kondo peaks to go down, until they
completely disappear. On the other hand, we found that the broad peak which 
corresponds to the intradot level essentially does not change in this range
of temperatures (not shown). These characters are in excellent agreement
with those of the LDOS. In fact, our investigations show that when 
$k_BT<\Gamma_1+\Gamma_3$, the excellent agreements between $dI_2/dV_2$
and LDOS are always maintained. 

To summarize, we have proposed and analyzed a possibility to experimentally
directly observe the local density of states of a quantum dot, thereby 
directly detect Kondo resonance peaks in it. In particular, using an
extra weak link to the quantum dot, we showed that curves of $dI_2/dV_2$ 
versus $V_2$ can map out perfectly the curves of LDOS versus energy, 
if experimental conditions are set in the correct range. We also provided
an analysis to these conditions and found them to be reasonable and
therefore should be realizable. Indeed, it will be extremely interesting
to experimentally test these predictions.

{\bf Acknowledgments:}
We gratefully acknowledge financial support from NSERC of Canada and FCAR of
Quebec.  

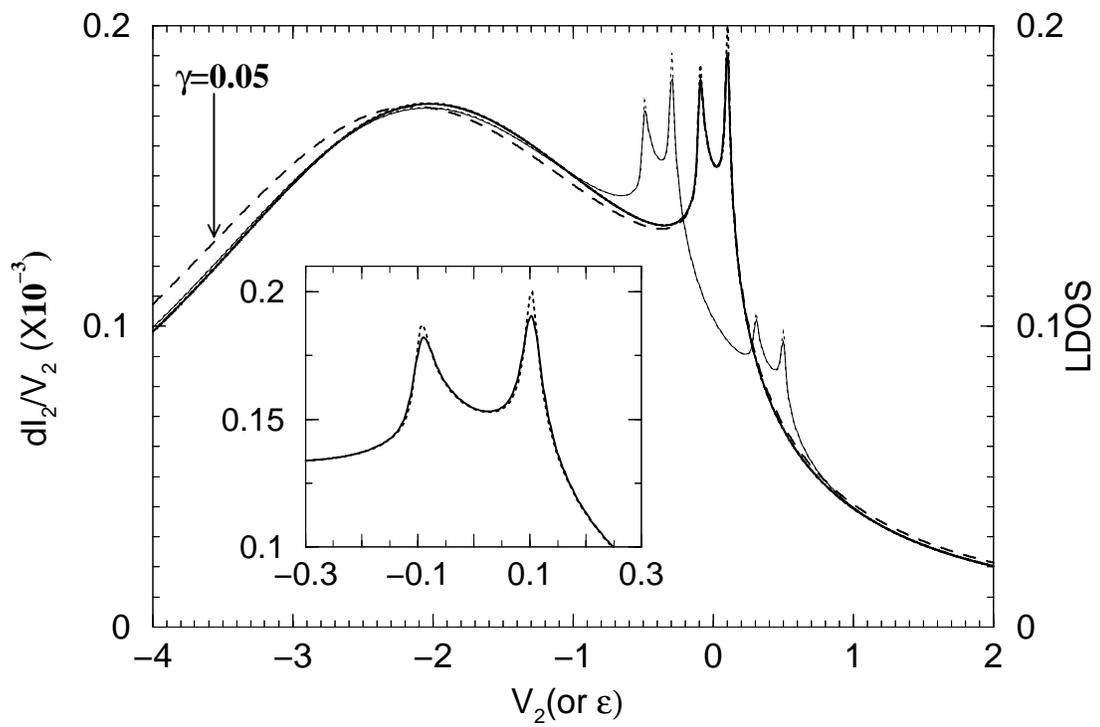
\begin{figure}
\caption{
Two solid curves are the differential conductance $dI_2/dV_2$ versus $V_2$, 
with $\Gamma_2=\Gamma_4=0.001$ and $\gamma=0$. Two dotted curves are 
the LDOS versus energy $\epsilon$ for the two-probe objective system 
(obtained by setting $\Gamma_2=\Gamma_4=0$). The thick dashed curve is 
$dI_2/dV_2$ versus $V_2$ with $\Gamma_2=\Gamma_4=0.001$ and $\gamma=0.05$. 
Other parameters are: $\Gamma_1=\Gamma_3=1$, $V_1=-V_3=0.1$, and 
$T=0.005$. The thick curves and thin curves correspond to
$\epsilon_{d\uparrow}(0)=\epsilon_{d\downarrow}(0)=-4.0$ and
$\epsilon_{d\uparrow}(0)=-4.2$, $\epsilon_{d\downarrow}(0)=-3.8$, 
respectively. Notice that the dotted curves almost overlap perfectly
with the solid curves so that they almost cannot be seen in the figure.
The inset shows the two Kondo resonance peaks at zero magnetic
field (by setting $\epsilon_{d\uparrow}(0)=\epsilon_{d\downarrow}(0)=-4.0$).
 }
\label{fig1}
\end{figure}

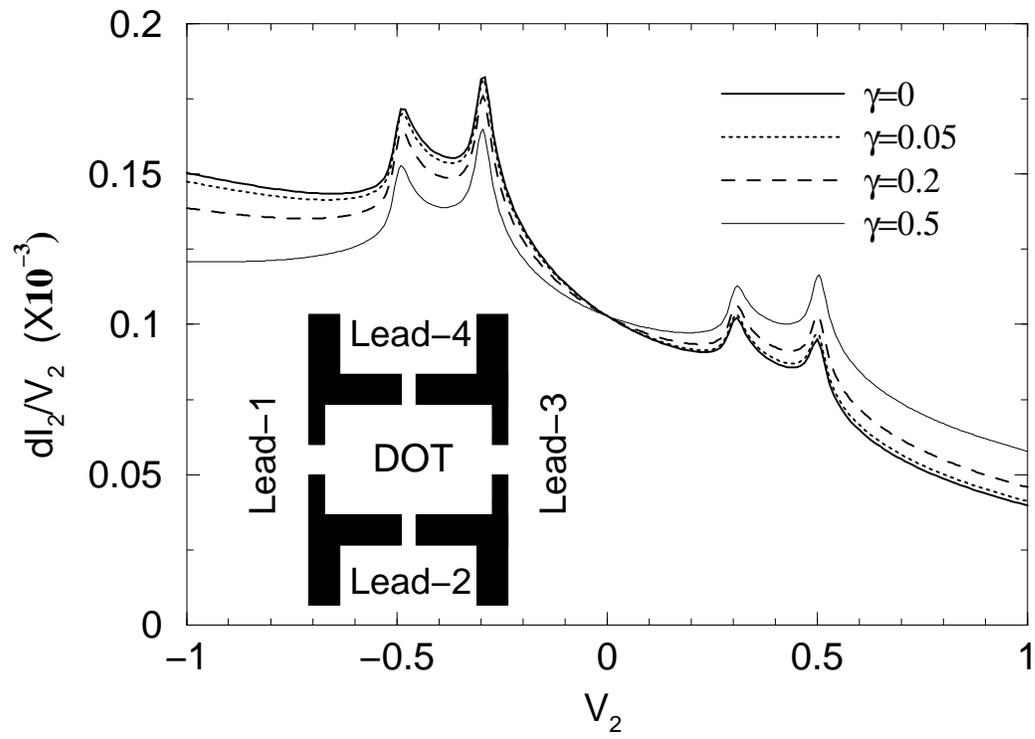
\begin{figure}
\caption{
$dI_2/dV_2$ versus $V_2$ at different parameters $\gamma$. Other 
parameters are same as those of the thin solid line of Fig.1. 
The inset is a schematic diagram for the four-probe quantum dot device.
}
\label{fig2}
\end{figure}

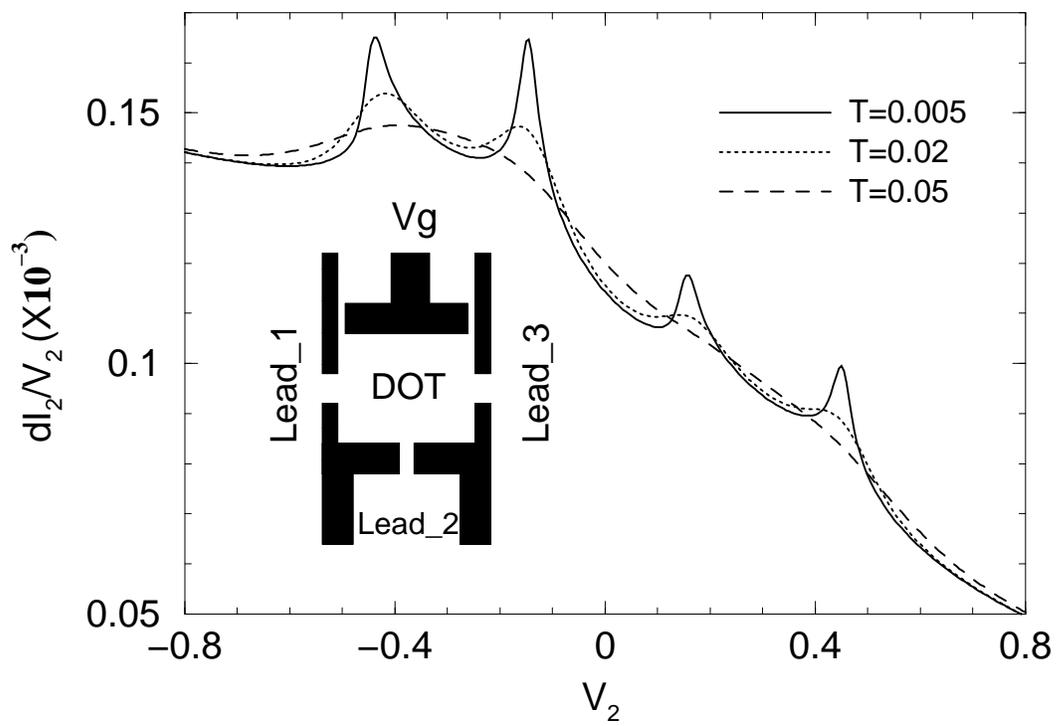
\begin{figure}
\caption{
$dI_2/dV_2$ versus $V_2$ for three different temperatures $T$, where 
the parameters are: $\Gamma_1=\Gamma_3=1$, $\Gamma_2=0.001$, $\gamma=0.05$, 
$\epsilon_{d\uparrow}(0)=-4.15$, $\epsilon_{d\downarrow}(0)=-3.85$, and 
$V_1=-V_3=0.15$. The inset is a schematic diagram for a proposed 
three-terminal quantum dot device.
}
\label{fig3}
\end{figure}


\begin{references} 
\bibitem{ref1}  
S. M. Cronenwett, T. H. Oosterkamp and L. P. Kouwenhoven, Science {\bf 281}, 540 (1998);
D. Goldhaber-Gordon, H. Shtrikman, D. Mahalu, D.A.-Magder, U. Meirav, and M.A.Kastner,
Nature {\bf 391}, 156 (1998);
T. Inoshita, Science {\bf 281}, 526 (1998).

\bibitem{ref2}
S. Sasaki, S. De Franceschi, J. M. Elzerman, W. G. van der Wiel,
M. Eto, S. Tarucha, and L. P. Kouwenhoven,
Nature {\bf 405}, 764 (2000);
W. G. van der Wiel, S. De Franceschi, T. Fujisawa, J. M. Elzerman, S. Tarucha, 
and L. P. Kouwenhoven, 
Science {\bf 289}, 2105 (2000).

\bibitem{ref3}
A. A. Clerk, V. Ambegaokar and S. Hershfield, 
Phys. Rev. B {\bf 61}, 3555 (2000).

\bibitem{ref4}
Q.-f. Sun, H. Guo and T.-h. Lin, cond-mat/0105120.

\bibitem{ref5}
V. Madhavan, W. Chen, T. Jamneala, M. F. Crommie, and N. S. Wingreen,
Science {\sl 280}, 567 (1998).

\bibitem{ref6}
J. Li, W.-D. Schneider, R. Berndt, and B. Delley, Phys. Rev. Lett. {\bf 80}, 2893
(1998).

\bibitem{ref7}
T. W. Odom, J.-L. Huang, C. L. Cheung, and C. M. Lieber, Science {\bf 290}, 1549 
(2000).

\bibitem{ref8}
W. Hofstetter, J. Konig and H. Schoeller, cond-mat/0104497.

\bibitem{ref9}
Y. Meir and N. S. Wingreen Phys. Rev. Lett. {\bf 68}, 2512 (1992).
 

\bibitem{ref11}
Y. Meir, N. S. Wingreen and P. A. Lee, Phys. Rev. Lett. {\bf 66}, 3048 (1991).

\bibitem{ref12}
The self-consistent equation is: 
$n_{\sigma}= Im \int \frac{d\epsilon}{2\pi}  G^<_{\sigma}(\epsilon)$.

\bibitem{ref13}
Y. Meir, N. S. Wingreen and P. A. Lee, Phys. Rev. Lett. {\bf 70}, 2601 (1993).


\bibitem{ref14}  
T.-K. Ng, Phys. Rev. Lett. {\bf 76}, 487 (1996).
 
\bibitem{ref15}
$LDOS(\epsilon)=-Im \sum_{\sigma}G_{\sigma}^r(\epsilon) /\pi$.

\bibitem{ref17}
R. Aguado and L. P. Kouwenhoven, Phys. Rev. Lett. {\bf 84}, 1986 (2000).



\end{references}
\end{document}